\documentclass[aps, prd, superscriptaddress, nofootinbib, preprintnumbers]{revtex4}
\usepackage{amsmath}
\usepackage{graphicx}

\newcommand{\chushi}[1]{}

\begin{document}

\title{Walking Technipions in a Holographic Model}       
\author{Masafumi Kurachi} \thanks{\tt kurachi@kmi.nagoya-u.ac.jp}
      \affiliation{ Kobayashi-Maskawa Institute for the Origin of Particles and the Universe (KMI) \\ 
Nagoya University, Nagoya 464-8602, Japan.}
\author{Shinya Matsuzaki}\thanks{\tt synya@hken.phys.nagoya-u.ac.jp}
      \affiliation{ Institute for Advanced Research, Nagoya University, Nagoya 464-8602, Japan.}
      \affiliation{ Department of Physics, Nagoya University, Nagoya 464-8602, Japan.}
\author{{Koichi Yamawaki}} \thanks{
      {\tt yamawaki@kmi.nagoya-u.ac.jp}}
      \affiliation{ Kobayashi-Maskawa Institute for the Origin of Particles and the Universe (KMI) \\ 
 Nagoya University, Nagoya 464-8602, Japan.}

%%%%%
\begin{abstract} 
We calculate 
masses of the technipions in 
the walking technicolor model with 
the anomalous dimension $\gamma_m =1$,  based on 
a holographic model which has a naturally light technidilaton $\phi$ as a composite Higgs with mass $m_\phi\simeq 125$ GeV. 
The one-family model (with 4 weak-doublets) 
is taken as a concrete example in such a framework, with the inputs being $F_\pi=v_{\rm}/2 \simeq 123$ GeV and $m_\phi\simeq 125$ GeV as well as $\gamma_m=1$. 
It is shown that technipion masses are 
enhanced by 
the large anomalous dimension to typically $O(1)$ TeV.
We find a correlation between the technipion masses  and  $S^{(\rm TC)}$, 
the $S$ parameter arising only from the technicolor sector. 
The current LHC data on the technipion mass limit thus constrains $S^{(\rm TC)}$ to be not as large as $O(1)$, giving a 
direct constraint on the technicolor model building. This 
is a new constraint on the technicolor sector alone quite independent of other sector connected by the extended-technicolor-type interactions,  in sharp contrast to the conventional $S$ parameter constraint from the precision electroweak measurements.
\end{abstract}

\maketitle

%%%%%
\section{Introduction} 
\label{sec:Introduction}
The mystery of the origin of the masses of the fundamental particles is one of the most important issues to be revealed in elementary particle physics. Even though the property of the 125 GeV scalar boson discovered at the LHC seems to be quite consistent with that of the Higgs boson in the Standard Model (SM) so far, there are several reasons for possible existence of the physics beyond the SM, 
such as the dynamical origin of the mass of the Higgs itself. 
Technicolor (TC)~\cite{Weinberg:1975gm,Yamawaki:1996vr} is an attractive candidate for such alternatives.  
Phenomenologically viable TC models, 
Walking TC (WTC) models~\cite{Yamawaki:1985zg,Bando:1986bg}, 
based on the 
approximately scale invariant dynamics 
(ladder Schwinger-Dyson (SD) equation), 
having a large mass anomalous dimension, $\gamma_m \simeq 1$,  predicted  
existence of a  
light composite scalar boson as a pseudo Nambu-Goldstone (NG) boson associated with the spontaneous breaking of the approximate scale invariance.  That is called the technidilaton, which can be identified with the 125 GeV boson discovered at the LHC~\cite{Matsuzaki:2012mk,Matsuzaki:2012vc},\cite{Matsuzaki:2012xx}. 

Recent lattice studies~\cite{Aoki:2013qxa}  actually indicate 
existence of a light flavor-singlet scalar bound state in the QCD with large number of massless flavors $N_f=8$, which is 
a candidate theory for the walking technicolor with anomalous dimension near unity as was suggested by several lattice results~\cite{Aoki:2013xza}. 
Such a light scalar could be a candidate for the technidilaton. (Note that a similar light scalar was also found on the lattice $N_f=12$ QCD~\cite{Aoki:2013zsa},
which may be a generic feature of the conformal dynamics, though not walking.)

Here we note that as we have repeatedly emphasized~\cite{Yamawaki:2010ms}\cite{Matsuzaki:2012mk,Matsuzaki:2012vc,Matsuzaki:2012xx},  the technidilaton cannot be exactly massless: Although the scale symmetry is {\it spontaneously broken} by the  condensate of the technifermion bilinear operator which is non-singlet under the scale transformation, it is at the same time
{\it explicitly broken}  {\it by the same condensate}, giving rise to the nonperturbative scale anomaly even if the coupling 
is non-running in the perturbative sense (The coupling runs by the condensate formation due to nonpertuabative dynamics).
Thus it is this {\it non-zero technidilaton mass} $M_\phi$, arising {\it from the noperturbative scale anomaly} due to the chiral condensate in WTC,
that can {\it dynamically explain the origin of the Higgs mass}
which is left mysterious in SM. This is somewhat {\it analogous to the $\eta^\prime$ meson} in the ordinary QCD, where the chiral $U(1)_A$ symmetry is 
spontaneously broken by the condensate and  {\it explicitly broken also by the anomaly} (the chiral  $U(1)_A$ anomaly) and hence cannot be exactly massless:
Nevertheless it is regarded as a pseudo NG boson a la Witten-Veneziano having a parametrically massless limit i.e., the large $N_c$ limit (with fixed ratio $N_f/N_c (\ll 1)$, Veneziano limit) in a way that  the chiral $U(1)_A$ anomaly tends to zero (only as a limit, not exactly zero).

The mass of the technidilaton  as a pseudo-NG boson comes from the nonperturbative trace anomaly $\theta_\mu^\mu \ne 0$ due to the chiral condensate and can be estimated 
through the partially conserved dilatation current (PCDC) relation~\cite{ 
Bando:1986bg}. 
A precise ladder evaluation of $M_\phi F_\phi$ based on this PCDC relation reads~\cite{Hashimoto:2010nw}: 
$(M_\phi F_\phi)^2 =- 4 \langle \theta_\mu^\mu \rangle \simeq  0.154 \cdot N_f N_{\rm TC}\cdot m_D^4 \simeq \left(2.5 \cdot v_{\rm EW}^2\right)^2 \cdot \left[(8/N_f)
(4/N_{\rm TC})\right]$, where $v_{\rm EW}^2=(246\,  {\rm GeV})^2 
= N_D F_\pi^2 \simeq 0.028 \cdot N_f N_{\rm TC} \cdot m_D^2$ (Pagels-Stokar formula),
with $N_D (=N_f/2)$ being the number of the electroweak doublets for $SU(N_{\rm TC})$ gauge theory. 
Thus the mass of the LHC Higgs,  $M_\phi \simeq
125\, {\rm GeV} \simeq v_{\rm EW}/2$, can be obtained, 
when we take $v_{\rm EW}/F_\phi =2F_\pi/F_\phi \simeq 1/5=0.2$  ($v_{\rm EW}= 2 F_\pi$ for $N_{\rm TC}=4, N_f=8$ (Farhi-Susskind one-family model~\cite{Farhi:1980xs}).  
Amazingly, this value of $F_\phi$ turned out to be
consistent with the LHC Higgs data~\cite{Matsuzaki:2012mk}.

Note the scaling $M_\phi/v_{\rm EW} \sim
1/\sqrt{N_f N_{\rm TC}}$ and  $F_\phi/v_{\rm EW} \sim (N_{\rm TC} N_f)^0$,  which is {\it a generic result independent of the ladder approximation}.
This implies existence of a limit, so-called Veneziano limit, namely the large $N_{\rm TC}$ limit with $N_f/N_{\rm TC}=$ fixed ($\gg 1$, though) (so as to be close to the conformal window), where the technidilaton has 
a vanishing mass $M_\phi/v_{\rm EW}\rightarrow 0$ in such a way that the nonperturbative scale anomaly in units of the weak scale 
vanishes $\langle \theta_\mu^\mu\rangle/v_{\rm EW}^4 \sim 1/(N_f N_{\rm TC}) \rightarrow 0$ in that limit. Thus a light technidilaton  $M_\phi/v_{\rm EW} \ll 1$ is naturally realized near such a Veneziano limit as is the case of the walking regime of the large-$N_f$ QCD~\cite{KMY}~\footnote{ 
One might think that such a large $N_f$ (and $N_{\rm TC}$) would result in the so-called $S$ parameter problem. 
We shall later discuss that it is not necessarily the case. }~\footnote{
More specific  computation~\cite{Harada:2003dc} 
via the ladder Bethe-Salpeter (BS) equation combined with the ladder SD equation in the large $N_f$ QCD 
 implies $M_\phi  \sim 4 F_\pi$ 
in the walking regime. Although it is much lighter than 
the techni-vector/axialvector
with mass $\sim 12 F_\pi$,  
it implies 
$M_\phi \simeq 500\, {\rm GeV}$~\cite{Yamawaki:2010ms} 
in the one-family model, which is still somewhat larger than the LHC Higgs. 
Such a ladder BS calculation as it stands corresponds to the flavor non-singlet scalar mass, totally ignoring the full gluodynamics such as the mixing with the glueball and effects of the axial $U(1)_A$ anomaly (instanton effects). Inclusion of such full gluodynamics 
will further lower the flavor-singlet scalar meson mass~\cite{KMY}. }.

More recently, in a holographic WTC model~\cite{Haba:2010hu} which we are based on this paper, it was shown~\cite{Matsuzaki:2012xx} that with 
the holographic parameter $G$ corresponding to the gluon condensate, we have 
$M_\phi/(4\pi F_\pi) \simeq 3/(2\sqrt{N_c}) (1+G)^{-1} \rightarrow 0$ with $F_\phi/F_\pi \simeq \sqrt{2N_f}$ as $G\rightarrow \infty$. This implies that 
the scale symmetry is parametrically realized in a way that the nonperturbative scale anomaly does vanish: 
$\langle \theta_\mu^\mu \rangle = - (M_\phi^2 F_\phi^2)/4 \sim v_{\rm EW}^4 /(N_f N_c) (1 +G)^{-2} \rightarrow 0$, 
in the strong gluon condensate limit $G \rightarrow \infty$ due to an additional factor $1/(1+G)^{-2}$ (besides the factor $1/(N_f N_c)$). 
It was also shown~\cite{Matsuzaki:2012xx}  that in this $G \rightarrow \infty$ 
limit the technidilaton behaves as a NG boson much lighter  than other bound sates such as the techni-$\rho$ and techni-$a_1$: $M_\phi/M_\rho, M_\phi/M_{a_1}
\rightarrow 0$. This is indeed analogous  to the flavor-singlet $\eta^\prime$ meson which parametrically behaves a NG boson of the axial $U(1)_A$ symmetry a la Witten-Veneziano in the large $N_c$ limit, as we mentioned above. 
In fact it naturally realizes $M_\phi \simeq F_\pi \simeq 125$ GeV for $G\simeq 10$ in the one-family model~\cite{Matsuzaki:2012xx}. 
Besides the lattice studies mentioned above, similar arguments for realizing such a parametrically light dilaton are given in somewhat 
different holographic contexts~\cite{Elander:2009pk}.

Once the technidilaton mass is tuned to be 125 GeV, the holography determines the technidilaton couplings (essentially controlled by the 
decay constant  $F_\phi$) to the SM particles, which nicely
reproduce the present LHC data for the 125 GeV Higgs, where the best fit value of the technidilaton decay constant  in the case of one-family model is
$v_{\rm EW}/F_\phi= 0.2-0.4  \,  (N_{\rm TC} =4,3)$ (depending on the electroweak singlet technifermion numbers)~\cite{Matsuzaki:2012xx}.

In this paper, based on the holographic model of Ref.~\cite{Haba:2010hu,Matsuzaki:2012xx}, 
we study another phenomenological issue of the generic WTC, the technipions, which are the left-over (pseudo) NG bosons besides the (fictitious) NG bosons absorbed into
SM gauge bosons. They exist in a large class of the WTC having large $N_f$, $N_f >2$ and will be a smoking gun of this class of WTC in the future LHC. 
 As a concrete realization of the WTC, we here consider the Farhi-Susskind one-family model~\cite{Farhi:1980xs}, inspired by the lattice studies 
  on $N_f=8$ QCD, which as already mentioned suggest  existence of a light flavor-singlet scalar~\cite{Aoki:2013qxa} 
and the walking behavior as well~\cite{Aoki:2013xza}. 
The model consists of $N_{\rm TC}$ copies of a whole generation of the SM fermions, 
in such a way that the TC sector of the model is $SU(N_{\rm TC})$ gauge theory with four weak-doublets, 
namely eight fundamental Dirac fermions $N_f=2 N_D=8$.  
The global chiral symmetry breaking pattern is then $SU(8)_L \times SU(8)_R/SU(8)_V$, 
resulting in the emergence of 63 NG bosons. 
Three of them are eaten by the SM weak gauge bosons, while 60  technipions remain as physical states. 
All the technipions become massive through 
the explicit breaking of the chiral $SU(8)_L \times SU(8)_R$ symmetry due to the  
SM gauge interactions and extended TC (ETC) gauge interactions,
and thereby become pseudo NG bosons.  
Estimation of the masses of these technipions are very important for the studies of collider phenomenologies, 
though it is a challenging task due to the need of non-perturbative calculations.  
This paper is the first attempt to compute the technipion masses using models of holographic dual. 

In Ref.~\cite{Harada:2005ru}, the mass of the charged technipions 
originating from the electromagnetic interaction and also the analogous mass of the colored technipions 
were studied in the large $N_f$ QCD by 
using the BS equation with the improved ladder approximation with two-loop running coupling. 
Further in Ref.~\cite{Jia:2012kd}, all the masses of technipions in the one-family model were estimated in the 
ladder analysis. 
However, it should be noted that the results obtained by using the improved ladder analysis, though qualitatively good,  have ambiguities in quantitative estimate which come from their systematic uncertainty originating from the approximation itself. Also, recent lattice result \cite{Aoki:2013xza} indicates that $SU(3)$ gauge theory with 8 fundamental 
fermions (which is exactly the case for the one-family TC with $N_{\rm TC}=3$) possesses walking nature, which is contrasted with 
the improved ladder analysis showing that it  
is deep inside the chiral symmetry breaking phase without walking signals. We would need other nonpertubative method having a different systematic uncertainty to make the quantitative estimate more reliable to be compared with the experiments.

Besides from-the-first-principle calculations on the lattice, one such a method would be the holographic approach which is based on the gauge-gravity duality~\cite{Maldacena:1997re} and is more useful for the phenomenological studies in the sense that desirable values of phenomenologically relevant quantities can be easily obtained by tuning input parameters of the effective holographic model. 
In the application to the WTC, 
based on the  popular  (bottom-up) holographic QCD~\cite{DaRold:2005zs, Erlich:2005qh}, 
with setting the anomalous dimension $\gamma_m=0$ through the bulk scalar mass parameter, we shall engineer 
 the walking theory by implementing  the  large anomalous dimension $\gamma_m=1$ 
instead of  $\gamma_m=0$~\cite{Hong:2006si,Haba:2008nz,Haba:2010hu,Matsuzaki:2012xx}.  
Particularly in the model having the technidilaton~\cite{Haba:2010hu,Matsuzaki:2012xx}, 
we can tune the input holographic parameters,  
besides the mass anomalous dimension, so as to adjust the physical quantities such as the mass of the composite Higgs (technidilaton), 
the weak scale, as well as the $S$ parameter to the experimental values. 
Once those holographic parameters are fixed, other phenomenological quantities which can be calculated from the holographic model become predictions. 
In Ref.~\cite{Kurachi:2013cha}, by using the same holographic model as Refs.~\cite{Haba:2010hu,Matsuzaki:2012xx}, the hadronic leading order contributions in QCD and WTC to the anomalous magnetic moment ($g-2$) of leptons were calculated. It was shown that, in the case of the real-life QCD, the known QCD contributions to $g-2$ of leptons are correctly reproduced, 
then it was applied to the calculation of the contribution from the 
WTC dynamics.

Here we adopt a similar approach for the estimation of the masses of 
the technipions in the one-family model of WTC, based on the first order perturbation of the explicit chiral symmetry breaking by the ``weak gauge couplings'' of SM gauge interactions and the ETC gauge interactions (Dashen's formula), while the full nonperturbative contributions of WTC sector are included by the holography (or its effective theory). This is the same strategy as the QCD estimate of the $\pi^+ - \pi^0$ mass difference, where the explicit chiral symmetry breaking is given by the QED lowest order coupling, while the full QCD nonperturbative contributions are estimated through the current correlators by various method like ladder, holography, lattice etc..   

It will be shown that technipion masses in the one-family model 
are enhanced due to the walking dynamics 
to the order of typically $O(1)$ TeV, qualitatively the same as  
the previous estimate~\cite{Jia:2012kd}, with  somewhat larger value. The large enhancement of the technipion mass has long been noted to be a generic feature of the large anomalous dimension~\cite{Holdom:1981rm}, and concretely shown in the explicit walking dynamics with $\gamma_m=1$ based on the ladder SD equation~\cite{Yamawaki:1985zg} and in the large $N_f$ QCD with $\gamma_m \simeq 1$~\cite{Harada:2005ru}. 
Striking fact is that 
although the explicit chiral symmetry breakings are formally
very small due to the ``weak gauge couplings'',  the nonperturbative contributions from the WTC sector lift all the technipions masses to the TeV region so that they all lose the nature of the ``pseudo NG bosons''.   
This is actually a universal feature of the dynamics with large anomalous dimension, ``amplification of the symmetry violation''~\cite{Yamawaki:1996vr},  as dramatically shown in the top quark condensate model~\cite{Miransky:1988xi}, 
based  on the Nambu-Jona-Lasinio model with large anomalous dimension $\gamma_m =2$. Note that although the left-over light spectrum are just three exact NG bosons absorbed into $W/Z$ bosons, 
our theory is completely  different from the $N_f=2$ model with the symmetry breaking of $SU(2)_L \times SU(2)_R/SU(2)_V$. In fact the three exact NG bosons as  well as the technidilaton 
are composite of the linear combinations of all the $N_f=8$ technifermions.

Here it is to be noted that the possible back reaction of the SM sector to the technidilaton mass through these weak gauge couplings
(top loop and EW gauge boson loops), which is
of potentially large quadratic divergence, actually was computed in the effective theory for WTC (dilaton chiral perturbation theory) coupled weakly to the SM sector, the result being  
negligibly small due to the largeness of $F_\phi$ (see Section III B of  Ref.~\cite{Matsuzaki:2012vc}).

We also show that there is a correlation between the technipion masses and $S^{({\rm TC})}$ which is the magnitude of the contribution to the $S$ parameter only from the TC sector,  
and that the latter cannot be as large as $O(1)$ due to the constraints from the currently available LHC data on the masses of the technipions. This is a new constraint on the TC dynamics alone, quite independently of
the conventional $S$ parameter constraint from the precision electroweak measurements which may involve not only the TC sector but the large contributions from other sector through the ETC interactions in such a way that they could largely cancel each other, as suggested 
in the Higgsless models \cite{Cacciapaglia:2004rb}.

This paper is organized as follows. In the next section, after the one-family model is briefly reviewed, 
the holographic model formulated in Refs.~\cite{Haba:2010hu,Matsuzaki:2012xx} is 
applied 
for the calculation of the masses of the technipions in the one-family model. 
Constraints from the currently available  LHC data, as well as implications for the future collider phenomenology are discussed in Sec.~\ref{sec:discussions}. Section \ref{sec:summary} is devoted to the summary of the paper. In Appendix~\ref{app:pion}, as a check of reliability of our calculations, we monitor 
the same holographic method by applying 
the estimation of 
the masses of colored technipions  
to 
the $\pi^+ - \pi^0$ mass difference in the real-life QCD. 
In Appendix~\ref{app:ladder}, the current correlator obtained from the holographic calculation is compared to that obtained from ladder BS calculation, the result being consistent each other.

%%%%%
\section{Holographic Estimate of Technipion Masses} 
%\label{sec:introduction}

In the Farhi-Susskind one-family model~\cite{Farhi:1980xs}, 
eight flavors of technifermions (techniquarks $Q_c$ and technileptons $L$) are introduced: $Q_c\equiv (U_c, D_c)^T$ 
(where $c=r,g,b$ is the QCD color charge) and $L \equiv (N, E)^T$, all having $SU(N_{\rm TC}$) charge, which are further embedded in a larger extended TC group, 
say $SU(N_{\rm TC}+3)$, by involving three generations of SM fermions. The chiral symmetry therefore is 
$SU(8)_L \times SU(8)_R$, which is broken by the technifermion condensation $\langle \bar{F}F \rangle \neq 0$ ($F=Q, L$) down to $SU(8)_V$, 
resulting in the emergence of 63 NG bosons. Three of them are eaten by $W$ and $Z$ bosons, while other 60 remain as physical states. 
Those are called technipions. Technipions obtain their masses through the explicit breaking effects 
(such as SM gauge interactions and extended TC four-fermion interactions), and become pseudo NG bosons. 
The technipions are classified  by the isospin and QCD color charges, which are listed in Table~\ref{tab:TP} together 
with the currents coupled to them, 
where the notation follows the original literature~\cite{Farhi:1980xs}. 
 For construction of the chiral Lagrangian described by those technipions, readers may refer to Ref.~\cite{Jia:2012kd}.

\begin{table}
\begin{tabular}{c|c|c|c}
\hline  \hline 
\hspace{15pt}
techni-pion 
\hspace{15pt}
&
\hspace{15pt} 
color 
\hspace{15pt}
&
\hspace{15pt} 
isospin 
\hspace{15pt}
& 
\hspace{15pt}
current 
\hspace{15pt}
\\  
\hline 
$\theta_a^{i}$ & octet & triplet & $\frac{1}{\sqrt{2}}\bar{Q} \gamma_\mu \gamma_5 \lambda_a \tau^{i} Q$ \\ 
$\theta_a$ & octet & singlet & $ \frac{1}{2\sqrt{2}} \bar{Q} \gamma_\mu \gamma_5 \lambda_a Q$ \\ 
$T_c^{i}$ $(\bar{T}_c^{i})$ & triplet & triplet & $\frac{1}{\sqrt{2}} \bar{Q}_c \gamma_\mu \gamma_5 \tau^{i} L$ (h.c.) \\ 
$T_c$ ($\bar{T}_c$) & triplet & singlet & $ \frac{1}{2 \sqrt{2}} \bar{Q}_c \gamma_\mu \gamma_5 L $ (h.c.)   \\ 
$P^i $ & singlet & triplet & $\frac{1}{2 \sqrt{3}} (\bar{Q} \gamma_\mu \gamma_5 \tau^i Q - 3 \bar{L} \gamma_\mu \gamma_5 \tau^i L)$ \\ 
$P^0$ &  singlet & singlet & $\frac{1}{4 \sqrt{3}} (\bar{Q} \gamma_\mu \gamma_5  Q - 3 \bar{L} \gamma_\mu \gamma_5  L)$  \\ 
\hline \hline 
\end{tabular} 
\caption{ The technipions and their color and isospin representation, as well as associated currents in the one-family model~\cite{Farhi:1980xs}. 
Here $\lambda_a$ ($a=1,\cdots,8$) are the Gell-Mann matrices, 
$\tau^i$ $SU(2)$ generators normalized as 
$\tau^i=\sigma^i/2$ ($i=1,2,3$) with the Pauli matrices $\sigma^i$, and 
the label $c$ stands for the QCD color index $c=r,g,b$. 
}
\label{tab:TP}
\end{table}

The holographic model proposed in Ref.~\cite{Haba:2010hu} is based on %deformations of 
a bottom-up approach for 
holographic dual of QCD (``hard-wall model'')~\cite{DaRold:2005zs,Erlich:2005qh}, with the input of the anomalous dimension $\gamma_m =0$ of QCD case being simply replaced by $\gamma_m =1$, the value expected in the walking theory. 
The model incorporates $SU(N_f)_L \times SU(N_f)_R$ gauge theory defined on the five-dimensional anti-de~Sitter (AdS) space-time, 
which is characterized by the metric $ds^2= g_{MN} dx^M dx^N = \left(L/z \right)^2\big(\eta_{\mu\nu}dx^\mu dx^\nu-dz^2\big)$ 
with $\eta_{\mu\nu}={\rm diag}[1, -1, -1,-1]$. 
Here, $M$ and $N$ ($\mu$ and $\nu$) represent five-dimensional (four-dimensional) Lorentz indices, 
and $L$ denotes the curvature radius of the AdS background. The fifth direction, denoted as $z$, is compactified on an interval, $ \epsilon \leq z \leq z_m  $, where $z=\epsilon$ (which will be taken to be 0 after all calculations are done) is the location of the ultraviolet (UV) brane while  $z=z_m$ is that of the infrared (IR) brane. The model introduces a bulk scalar $\Phi_S$ which transforms as a bifundamental representation field under the $SU(N_{f})_L \times SU(N_{f})_R$ gauge symmetry; a field which is dual to the 
(techni-)fermion bilinear operator $\bar{F} F$. 
The mass parameter for $\Phi_S$, $m_{\Phi_S}$, is thus holographically related to $\gamma_m$ as 
$ 
m_{\Phi_S}^2=- (3-\gamma_m)(1+ \gamma_m)/L^2 
$. 
When we apply the model for the calculations of physical quantities in WTC models, we take $\gamma_m = 1$. 
In addition, as in Ref.~\cite{Haba:2010hu,Matsuzaki:2012xx} we include another bulk scalar, $\Phi_G$, dual to the (techni-)gluon condensation operator $ G_{\mu\nu}^2$, which has vanishing 
mass parameter since its conformal dimension is taken to be 4~\footnote{
Note that in our holographic model based on the popular static hard wall model~\cite{DaRold:2005zs,Erlich:2005qh}, having the IR brane at $z_m$ fixed by hand 
(which explicitly breaks the conformal invariance in the five-dimension), 
the dilaton/radion for stabilizing the IR brane at $z_m$ as that discussed in some holographic models~\cite{Rattazzi:2000hs}  is set to have a large mass of order ${\mathcal O}(1/z_m)
=$ several TeV's (See Table \ref{para-set}) 
and is irrelevant to our discussions of the technidilaton.
In fact our technidilaton~\cite{Haba:2010hu,Matsuzaki:2012xx}   
is identified as a bound state of technifermion and anti-technifermion, which holographically corresponds to  
the ground state in Kaluza-Klein (KK) modes for the flavor-singlet part of the bulk scalar $\Phi_S$, 
 in sharp contrast to the radion and dilaton in other holographic WTC models\cite{Elander:2009pk,Jarvinen:2011qe}.
Actually, the flavor-singlet part in  $\Phi_S$ mixes with a glueball-like scalar from $\Phi_G$. However, as shown in Ref.~\cite{Haba:2010hu}, 
the mixing turns out to be negligible when one requires the present holographic model to reproduce the UV asymptotic behaviors of current correlators
in OPE. Moreover, the lowest glueball as the lowest KK mode of $\Phi_G$ was explicitly computed to be near 20 TeV, much heavier than the 125 GeV technidilaton as the fermionic bound state from $\Phi_S$ in the walking case
($\gamma_m=1,\, G \simeq 10$)~\cite{Matsuzaki:2012xx}, which is  in sharp 
contrast to the QCD case ($\gamma_m =0, \,G\simeq 0.25$) where both glueball (from $\Phi_G$) and the flavor-singlet fermionic bound state (from $\Phi_S$) are 
comparably heavy $\simeq 1.2-1.3$ GeV (and may be strongly mixed)~\cite{Kurachi:2013cha}.
}. 
Thanks to this additional explicit bulk scalar field $\Phi_G$, this holographic model is the only model which naturally improves the matching with the OPE of the underlying theory (QCD and WTC) for current correlators so as to reproduce gluonic $1/Q^4$ term. This term is clearly distinguished from the same $1/Q^4$ term from chiral condensate in the case of WTC with $\gamma_m=1$ 

The action of the model is given as~\cite{Haba:2010hu} 
\begin{equation} 
  S_5 = S_{\rm bulk} + S_{\rm UV} + S_{\rm IR} 
  \,, \label{S5}
\end{equation}
where $S_{\rm bulk}$ denotes the five-dimensional bulk action, 
\begin{eqnarray} 
  S_{\rm bulk} 
  &=& 
  \int d^4 x \int_\epsilon^{z_m} dz 
  \sqrt{g} 
  \frac{1}{g_5^2} \, e^{c_G g_5^2 \Phi_G} 
 \Bigg[ 
\frac{1}{2} \partial_M \Phi_G \partial^M \Phi_G 
\nonumber \\ 
&& 
+ {\rm Tr}[D_M \Phi_S^\dag D^M \Phi_S - m_{\Phi_S}^2 \Phi_S^\dag \Phi_S ] 
\nonumber \\ 
&&
  - \frac{1}{4} {\rm Tr}[L_{MN}L^{MN} + R_{MN}R^{MN}] 
 \Bigg] 
 \,, \label{S:bulk}
\end{eqnarray}
and $S_{\rm UV, IR}$ are the boundary actions which are given in Ref.~\cite{Matsuzaki:2012xx}.   
The covariant derivative acting on $\Phi_S$ in Eq.(\ref{S:bulk}) 
 is defined as $D_M\Phi_S=\partial_M \Phi_S+iL_M\Phi_S-i\Phi_S R_M$, where 
$L_M(R_M)\equiv L_M^a(R_M^a) T^a$ with $L_M (R_M)$ being the five-dimensional gauge fields and $T^a$ being the generators of 
$SU(N_{f})$ which are normalized as 
${\rm Tr}[T^a T^b]=\delta^{ab}$. 
$L(R)_{MN}$ is the five-dimensional field strength which is defined as 
$L(R)_{MN} = \partial_M L(R)_N - \partial_N L(R)_M 
 - i [ L(R)_M, L(R)_N ]$, and 
$g$ is defined as $g={\rm det}[g_{MN}]= (L/z)^{10}$.

The five-dimensional vector and axial-vector gauge fields $V_M$ and $A_M$ are defined as 
$ V_M = (L_M + R_M)/\sqrt{2}$ and 
$A_M = (L_M-R_M)/\sqrt{2}$. 
It is convenient to work with the gauge-fixing $V_z=A_z\equiv 0$ and take the boundary conditions 
$V_\mu(x,\epsilon)=v_\mu(x)$, $A_\mu(x,\epsilon)=a_\mu(x)$ and  $\partial_z V_\mu(x,z)|_{z=z_m}=\partial_z A_\mu(x,z)|_{z=z_m}= 0$, 
where $v_\mu(x)$  and $a_\mu(x)$ correspond to sources for the vector and axial-vector currents, respectively. 
We solve the equations of motion for (the transversely polarized components of) the Fourier transformed fields,  
$V_\mu(q,z)=v_\mu(q) V(q,z)$ and $A_\mu(x,z)=a_\mu(q) A(q,z)$, where 
$V(q,z)$ and $A(q,z)$ denotes the profile functions for the bulk vector and axial-vector gauge fields.

We then substitute the solutions back into the action in Eq.(\ref{S:bulk}), to obtain the 
generating functional $W[v_\mu, a_\mu]$ holographically dual to WTC. 
Evaluating the UV asymptotic behaviors of the vector and axial-vector current correlators $\Pi_V(Q^2)$ and $\Pi_A(Q^2)$ 
with $Q^2\equiv -q^2$, 
we can thus fix 
the gauge coupling $g_5$ and the parameter $c_G$ appearing in the action 
to match the asymptotic forms with the expressions expected from the operator product expansion (OPE):~\cite{Haba:2010hu} 
\begin{equation} 
\frac{ L}{g_5^2} = \frac{N_{\rm TC}}{12\pi^2}
 \,, \qquad 
c_G = - \frac{N_{\rm TC}}{192 \pi^3}  
\,, 
\end{equation}

After the value of $\gamma_m$ is taken to be $1$, and a specific number for $N_{\rm TC}$ is fixed, 
remaining parameters in the holographic model are $\xi$, $z_m$, and $G$, where 
$\xi$ and $G$ parametrize the IR values for the vacuum expectation values of the bulk scalars $\Phi_S$ and $\Phi_G$, $v_S(z=z_m)$ 
and $v_{\chi_G}(z=z_m)$~\cite{Haba:2010hu}: 
\begin{eqnarray} 
v_S(z_m) &=& \frac{\xi}{L}
\,, \nonumber \\ 
v_{\chi_G}(z_m)  &=& 1 + G 
\,, 
\end{eqnarray}
with $v_{\chi_G} = \langle \chi_G(x,z) \rangle \equiv \langle e^{c_G g_5^2\Phi_G(x,z)/2 } \rangle$. 
Hence, once three physical quantities are chosen to fix these parameters, we calculate all other quantities related to WTC models. 
We shall choose the technipion decay constant, $F_\pi$, the technidilaton mass, $M_\phi$, and the $S$ parameter (actually 
$S^{({\rm TC})}$ coming from only the TC sector, but denoted as $S$ hereafter) as those three. 
$F_\pi$ is taken to be $F_\pi=123$ GeV so that it reproduces the electroweak (EW) scale $v_{\rm EW}^2 = N_D F_\pi^2 = (246\ {\rm GeV})^2$ with $N_D=4$, where $N_D$ is the number of the EW doublets exist in the model.  
The technidilaton mass $M_\phi$ is taken to be $M_\phi=125$ GeV  to be identified with the LHC Higgs boson~\cite{Matsuzaki:2012xx}. 
As for the $S$ parameter, we take several values, namely $S=(0.1,0.3,1.0)$ for our study. 
This is because, although $S=0.1$ is a phenomenologically viable benchmark value, 
there is a possibility that even if the WTC dynamics itself produces a somewhat large value of $S$, contributions coming from other part of the model 
(such as the extended TC interactions) could partially cancel it in a way similar to the concept of 
fermion-delocalization effect studied in Higgsless models~\cite{Cacciapaglia:2004rb}. 
The values of parameters ($\xi$, $z_m$, 
$G$) which reproduce the above mentioned three physical quantities ($F_\pi, M_\phi, S$) for the cases of $N_{\rm TC}=3,4,$ and $5$ 
are summarized in Table~\ref{para-set}.
\begin{table}[t]
\begin{center}
\begin{tabular}{c|c|c|c}
\hline \hline 
\hspace{35pt} 
$N_{\rm TC}$ 
\hspace{35pt} 
& 
\hspace{35pt} 
$\xi$ 
\hspace{35pt} 
&  
\hspace{35pt} 
$G$  
\hspace{35pt} 
& 
\hspace{35pt} 
$z_m^{-1}[{\rm TeV}]$ 
\hspace{35pt} 
\\ 
 \hline 
3 
& 
(0.014, 0.024, 0.044) 
& 
(9.7, 9.7, 9.8) 
& 
(5.2, 3.0, 1.6) 
\\
4 & (0.015, 0.027, 0.048) & (8.3, 8.3, 8.4) & (4.7, 2.7, 1.5) 
\\ 
5 &  (0.016, 0.029, 0.052) & (7.3, 7.3, 7.4) & (4.4, 2.5, 1.4) 
\\
\hline \hline  
\end{tabular}
\end{center}
\caption{Parameter sets which reproduce $F_\pi=123$ GeV, $M_\phi=125$ GeV, and $S=(0.1,\, 0.3,\, 1.0)$}
\label{para-set}
\end{table}
In the following subsections, we estimate the masses of the technipions with the parameter sets listed there.

%%%%%
\subsection{Color-singlet technipion masses} 

The color-singlet technipions $P^0 \sim (\bar{Q} \gamma_5 Q - 3 \bar{L} \gamma_5 L)$ and 
$P^{i=1,2,3} \sim ( \bar{Q} \gamma_5 \sigma^i Q - 3 \bar{L} \gamma_5 \sigma^i L)$ listed in Table~\ref{tab:TP} 
obtain their masses through the extended TC-induced four-fermion interaction, 
\begin{equation} 
 {\cal L}_{\rm 4-fermi}^{\rm ETC} (\Lambda_{\rm ETC}) 
 = \frac{1}{\Lambda_{\rm ETC}^2} 
 \left(
 \bar{Q} Q \bar{L}L - \bar{Q} \gamma_5 \sigma^i Q \bar{L} \gamma_5 \sigma^i L
 \right)
 \,. 
\end{equation}
The masses can be estimated by using the current algebra as 
\begin{equation} 
 m_{P^{i,0}}^2 
 = \frac{1}{F_\pi^2} 
 \langle 0 | 
 [{\bf Q}_{P^{i,0}}, [{\bf Q}_{P^{i,0}}, {\cal L}_{\rm 4-fermi}^{\rm ETC}(\Lambda_{\rm ETC})]] 
 |0 \rangle  
 \,, 
\end{equation}
where ${\bf Q}_{P^{i,0}}$ denotes the chiral charges defined as ${\bf Q}_{P^{i,0}}= \int d^3 x \, J^0_{P^{i,0}}(x)$ with the corresponding 
currents $J^\mu_{P^{0,i}}(x)$ listed in Table~\ref{tab:TP}.  
 The $P^i$ and $P^0$ masses are thus evaluated to be~\cite{Jia:2012kd} 
\begin{eqnarray} 
 m_{P^0}^2 &=& \frac{5}{2} \frac{\langle 0| \bar{F}F |0 \rangle^2_{\Lambda_{\rm ETC}}}{F_\pi^2 \Lambda_{\rm ETC}^2}
 \,, \nonumber \\ 
  m_{P^i}^2 &=& 4 \frac{\langle 0| \bar{F}F |0 \rangle^2_{\Lambda_{\rm ETC}}}{F_\pi^2 \Lambda_{\rm ETC}^2}
\,, \label{mP}
\end{eqnarray}
where we used $\langle 0| \bar{L}L |0 \rangle=1/3 \langle 0| \bar{Q}Q |0 \rangle \equiv \langle 0| \bar{F}F |0 \rangle$. 
The ratio $m_{P^0}^2/m_{P^i}^2 =5/8$ is a salient prediction of the one-family model, independently of the walking
dynamics.

\begin{table} 
\begin{tabular}{c|c|c} 
\hline \hline 
\hspace{35pt} 
$N_{\rm TC} $
\hspace{35pt} 
& 
\hspace{35pt} 
$m_{P^0} [{\rm TeV}] $
\hspace{35pt} 
& 
 \hspace{35pt} 
$m_{P^i} [{\rm TeV}] $
\hspace{35pt}  
 \\ 
\hline 
3 & (2.3, 1.3, 0.72) & (2.9, 1.7, 0.91) 
\\
4 & (2.4, 1.4, 0.76) & (3.0, 1.7, 0.96)
\\
5 & (2.5, 1.4, 0.79) & (3.1, 1.8, 0.99) 
\\
\hline \hline 
\end{tabular} 
\caption{
The predicted values of the color-singlet technipion masses for 
$N_{\rm TC}=3,4$ and 5 with $F_\pi=123$ GeV, $M_\phi=125$ GeV and $S=(0.1,0.3,1.0)$ fixed. 
}
\label{tab:mP:vals}
\end{table}

The present holographic model  
gives a formula for the technifermion condensate $\langle 0| \bar{F}F |0 \rangle$ renormalized at the extended TC scale $\Lambda_{\rm ETC}$ 
as~\cite{Haba:2010hu,Matsuzaki:2012xx} 
\begin{equation} 
 \langle 0| \bar{F}F |0 \rangle_{\Lambda_{\rm ETC}} 
 = - \frac{\sqrt{3}N_{\rm TC}}{12\pi^2} \frac{\Lambda_{\rm ETC} \xi(1+G)}{z_m^2}
\,. 
\end{equation}
This allows us to express the $P^{i,0}$ masses in Eq.(\ref{mP}) as follows: 
\begin{eqnarray} 
 m_{P^0} &=& \sqrt{\frac{5}{2}} \frac{\sqrt{3} N_{\rm TC}}{12 \pi^2 F_\pi} \frac{\xi(1+G)}{ z_m^2} 
 \,, \nonumber \\ 
  m_{P^i} &=& 2  \frac{\sqrt{3} N_{\rm TC}}{12 \pi^2 F_\pi } \frac{\xi(1+G)}{z_m^2} 
\,. \label{mP:re}
\end{eqnarray} 
Using the parameter set given in Table~\ref{para-set}, we thus calculate the $P^{i,0}$ masses to obtain the numbers listed 
in Table~\ref{tab:mP:vals}~\footnote{
These $P^{0,i}$ mass values are somewhat larger than those obtained in Ref.~\cite{Jia:2012kd} based on 
estimate with help of the (improved) ladder SD  
analysis. 
This is because of the difference of the size of intrinsic mass scale obtained in both approaches: 
In the present holographic model, a typical hadron mass scale such as dynamical mass of technifermions $m_F$ 
is predicted to be $\simeq 4 \pi F_\pi={\cal O}({\rm TeV})$, while $m_F$ estimated by the ladder approximation tends to 
get smaller than 1 TeV. The larger $m_F$ gives rise to the larger size of hadron spectrum, except the technidilaton 
protected by the scale symmetry~\cite{Matsuzaki:2012xx}. 
}. 

With smaller $\xi$ values as in Table~\ref{para-set}, which ensures the presence of the light technidilaton~\cite{Matsuzaki:2012xx}, 
the technipion decay constant $F_\pi$ can be approximated to be 
\begin{equation} 
 F_\pi \simeq \sqrt{\frac{N_{\rm TC}}{12\pi^2}} \frac{\xi (1+G)}{z_m}
\,. 
\end{equation}
 Putting this into Eq.(\ref{mP:re}) thus leads to the approximate formula for the $P^{i,0}$ masses: 
\begin{eqnarray} 
 m_{P^0} &\simeq & \sqrt{\frac{5}{2}} \frac{\sqrt{N_{\rm TC}}}{2\pi z_m}
 \,, \nonumber \\ 
  m_{P^i} &\simeq & 2 \frac{\sqrt{N_{\rm TC}}}{2\pi z_m}
\,, \label{mP:approx} 
\end{eqnarray} 
by which one can check that the numbers listed in Table~\ref{tab:mP:vals} are well reproduced.

%%%%%
\subsection{Color-triplet and -octet technipion masses} 
\label{subsec:coloredTP}

The color-octet and -triplet technipions $\theta_a^{i(0)} \sim \bar{Q} \gamma_5 \lambda_a \sigma^{i}({\bf 1}_{2\times 2}) Q$ 
and $T_c^{i(0)} \sim \bar{Q}_c \gamma_5 \sigma^i ({\bf 1}_{2 \times 2})L$ listed in Table~\ref{tab:TP}
acquire the masses by QCD gluon interactions~\footnote{This is analogous to the electromagnetic effects on the QCD pion. 
In Appendix~\ref{app:pion}, we apply the same method used here for the calculation of $\pi^+ - \pi^0$ mass difference, 
and show that holographic calculation reproduce the experimental value to a good accuracy.}. 
 The masses can be estimated by assuming the one-gluon exchange contribution, given as an integration over the momentum carried by 
the vector and axial-vector correlators $\Pi_{V,A}$:     
\begin{equation} 
 m_{3,8}^2 = \frac{3 C_{3,8}}{4\pi F_\pi^2} \int_0^\infty dQ^2 \, \alpha_s(Q^2) \left[ \Pi_V(Q^2) - \Pi_A(Q^2) \right] 
\,, 
\label{m38eq}
\end{equation} 
with the group factor $C_{3(8)}=4/3(3)$ for the color-triplet (octet) technipion, 
$Q\equiv \sqrt{-q^2}$ being the Euclidean momentum. 
Again the ratio $m_3/m_8=4/9$ is a salient prediction of the one-family model, independently of the walking dynamics.  
In Eq.(\ref{m38eq}) we have incorporated the $Q^2$-dependence of the QCD gauge coupling $\alpha_s$. 

\begin{table} 
 \begin{tabular}{c|c|c} 
\hline \hline 
\hspace{35pt} 
$ N_{\rm TC} $ 
\hspace{35pt} 
& 
 \hspace{35pt} 
$ m_{3} [{\rm TeV}] $ 
\hspace{35pt} 
& 
\hspace{35pt} 
$ m_8 [{\rm TeV}] $ 
\hspace{35pt} 
\\ 
\hline 
 3  & (3.0, 1.9, 1.1) &  (4.6, 2.8, 1.6) 
\\ 
  4  & (2.9, 1.8, 1.0) &  (4.4, 2.7, 1.5) 
\\ 
 5 & (2.9, 1.7, 1.0) &  (4.3, 2.6, 1.5) 
\\ 
\hline \hline 
\end{tabular} 
\caption{ 
The predicted values of the color-triplet ($m_3$) and -octet ($m_8$) technipion masses for 
$N_{\rm TC}=3,4$ and 5 with $F_\pi=123$ GeV, $M_\phi=125$ GeV and $S=(0.1,0.3,1.0)$ fixed. 
}
\label{tab:m38:vals}
\end{table}

The present holographic model gives the formulas for 
$\Pi_{V}(Q^2)$ and $\Pi_{A}(Q^2)$ as~\cite{Haba:2010hu,Matsuzaki:2012xx} 
\begin{eqnarray}
  \Pi_{V(A)}(Q^2) = \frac{N_{\rm TC}}{12\pi^2} \frac{\partial_z V(A)(Q^2, z)}{z} \Bigg|_{z=\epsilon \to 0}
  \,. \label{PiVA}
\end{eqnarray} 
The vector and axial-vector profile functions $V(Q^2,z)$ and $A(Q^2,z)$ are determined by solving 
the following equations:   
\begin{eqnarray} 
&& 
\left[ 
 - Q^2 + \omega^{-1}(z) \partial_z \omega(z) \partial_z  
\right] V(Q^2, z) = 0 
\,, \nonumber \\ 
&& 
\left[ 
 - Q^2 + \omega^{-1}(z) \partial_z \omega(z) \partial_z  - 2 \left( \frac{L}{z} \right)^2 [v_S(z)]^2
\right]A(Q^2, z) = 0 
\,,
\label{EOM:PiA}
\end{eqnarray}
with 
the boundary conditions 
$V(Q^2,z)|_{z=\epsilon \rightarrow 0}=A(Q^2,z)|_{z=\epsilon \rightarrow 0}=1$. 
In Eq.(\ref{EOM:PiA}) the vacuum expectation value $v_S(z)$ and a function $\omega(z)$ are given as~\cite{Haba:2010hu,Matsuzaki:2012xx}    
\begin{eqnarray} 
v_S(z) &=& \frac{\xi(1+G)}{L} \frac{(z/z_m)^2}{1+G(z/z_m)^4} 
\frac{\log(z/\epsilon)}{\log(z_m/\epsilon)} 
\,, \nonumber \\ 
\omega(z) &=&\frac{L}{z} \left( 1 + G \left( \frac{z}{z_m} \right)^4 \right)^2 
\,. 
\end{eqnarray} 
Thus $(\Pi_V -\Pi_A)$ in Eq.(\ref{PiVA}) is evaluated as 
a function of the holographic parameters, $\xi, G$  
and the IR position $z_m$. 
(For details, see Refs.~\cite{Haba:2010hu,Matsuzaki:2012xx}.)

Using the parameter sets given in Table~\ref{para-set}, 
we thus estimate the colored technipion masses for $S=(0.1, 0.3, 1.0)$ to obtain the values given 
in Table~\ref{tab:m38:vals}~\footnote{
In Ref.~\cite{Jia:2012kd} the colored technipion masses were estimated to be somewhat smaller than those listed in Table~\ref{tab:m38:vals}. 
The estimate in Ref.~\cite{Jia:2012kd} was based on an assumption that 
$(\Pi_V-\Pi_A)$ in the integrand in Eq.(\ref{m38eq}) is dominated in the UV region and 
hence can be replaced with the OPE expressions from the UV cutoff down to some IR scale $\sim 4 \pi F_\pi$. 
As shown in the present study, however, such an assumption results in underestimate of the masses. 
This is the main reason for the difference between the estimated size of colored technipion 
masses in the present study and Ref.~\cite{Jia:2012kd}.  Apart from this point, the present holographic estimation of the mass of all the technipions
is  roughly consistent with the ladder estimate of Ref.~\cite{Jia:2012kd}, as long as compared with the parameter choice corresponding to the ladder
estimate.
}.   
In evaluating the integral over $Q^2$ in Eq.(\ref{m38eq})  
we have introduced the UV cutoff 
$\Lambda_{\rm UV}^2 = (4\times 10^6\ {\rm GeV})^2$, which is 
consistent with 
the  current bound from flavor changing neutral current~\cite{Chivukula:2010tn}. 
As for the QCD gauge coupling $\alpha_s(Q^2)$, the one-loop running coupling is used with taking 5, 6, and $(6+2N_{\rm TC})$ flavors as number of active colored fermions in regions $Q^2 < m_t^2$, $m_t^2 < Q^2 < (4\pi F_\pi)^2$, and $(4\pi F_\pi)^2 < Q^2$, respectively, where $(4\pi F_\pi)$ corresponds to the size of 
the dynamical mass scale of the technifermions estimated from the present holographic model~\cite{Haba:2010hu,Matsuzaki:2012xx}.   
The value of the coupling at the $Z$ boson mass scale, $\alpha_s(m_Z^2)=0.1182$~\cite{Beringer:1900zz}, is used as input, and an infrared regularization is introduced in such away that $\alpha_s(Q^2)$ takes constant value $\alpha_s(Q^2=1\, {\rm GeV}^2)$ in the region $Q^2 < 1\, {\rm GeV}^2$. We have checked that infrared regularization dependence is negligible for the estimation of technipion masses.

\begin{figure}[t] 
\begin{center} 
\includegraphics[width=9cm]{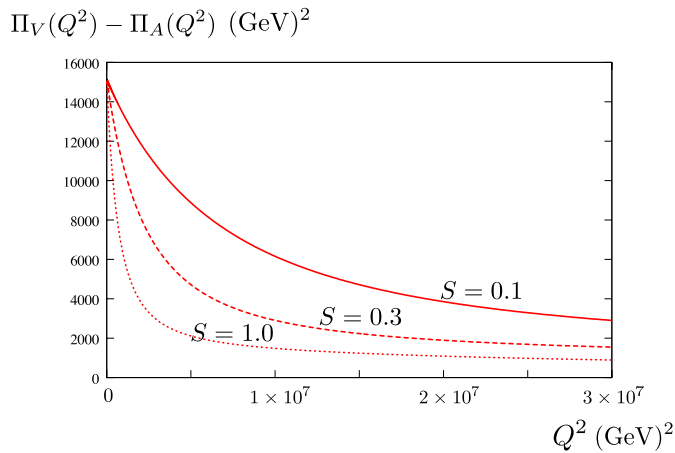} 
\end{center}
\caption{$\Pi_V(Q^2)-\Pi_A(Q^2)$ calculated from the holographic WTC model for the case of $N_{\rm TC} = 4$. Solid, dashed, and dotted curves correspond to the cases of $S=0.1$, $0.3$, and $1.0$, respectively.}  
\label{fig:WTC}
\end{figure}

In Fig.~\ref{fig:WTC}, we show $(\Pi_V(Q^2)-\Pi_A(Q^2))$ calculated from Eq.(\ref{PiVA}) 
for the case of $N_{\rm TC} = 4$. 
General tendencies of $(\Pi_V(Q^2)-\Pi_A(Q^2))$ for the case of  $N_{\rm TC} = 3$ and $5$ are similar 
to the case of $N_{\rm TC} = 4$.\footnote{In Appendix~\ref{app:ladder}, the current correlator obtained from the holographic calculation is compared to that obtained from ladder BS calculation.} In the figure, solid, dashed, and dotted curves correspond to the cases of $S=0.1$, $0.3$, and $1.0$, respectively. 
Since $(\Pi_V(Q^2)-\Pi_A(Q^2))$ at $Q^2=0$ is equal to $F_\pi^2$, which is adjusted to be $(123\ {\rm GeV})^2$ by tuning parameters 
so that the model reproduces $v_{\rm EW} = 2 F_\pi = 246$ GeV, 
all the curves take the same value at $Q^2=0$.  
However, slopes of each curve at $Q^2=0$ are different since it is proportional to the magnitude of the $S$ parameter, 
\begin{equation} 
 S = - 4 \pi N_D \frac{d}{d Q^2} (\Pi_V(Q^2) - \Pi_A(Q^2)) \Bigg|_{Q^2 =0}
 \,.  
\end{equation} 
From this, we understand the general tendency of the above results, namely the smaller the value of the $S$ parameter, heavier the masses of technipions due to the larger contribution in the integral of Eq.~(\ref{m38eq}). 
This is telling us the important thing. As we wrote earlier in this section, the EW precision measurements do not necessarily require that the contribution to the $S$ parameter from the TC dynamical sector, denoted as $S^{({\rm TC})}$ before,  is small, because the ultimate value of the $S$ parameter depends on how the TC sector is embedded in the whole model together with the SM fields. On the other hand, the correlation between the slope of $(\Pi_V(Q^2)-\Pi_A(Q^2))$ at $Q^2=0$ and the masss of colored technipions shown here rather directly constrain the contribution to the $S$ parameter from the TC dynamics itself if the experimental lower bound of the mass of the colored technipion become larger. In the next section, we will show that the current LHC bounds on the masses of technipions are already placing constraint that the 
contribution to the $S$ parameter from the TC sector
should no be as large as $O(1)$.

To summarize, we have calculated the masses of the technipions in the one-family TC model, and shown that those are rather heavy, raging from $\sim 1$ TeV to $4.6$ TeV depending on the $N_{\rm TC}$ and the input value of the $S$ parameter. One thing which should be noted here is that the existence of heavy technipions does not necessarily mean that the scale of explicit breaking of chiral symmetry is also large. Indeed, the reason why we obtained large technipion masses here is that the contribution from the UV energy scale is enhanced due to the large mass anomalous dimension $\gamma_m = 1$ though the explicit breaking itself is rather small perturbation. And accordingly, we expect the walking behavior persists all the way down to the intrinsic dynamical scale of the TC model, without decoupling of any degree of freedom of the technifermion.

%%%%%
\section{LHC phenomenology}
\label{sec:discussions}

As we have estimated  
the masses of technipions in the one-family model, we discuss collider phenomenologies related to the technipions in this section. First we discuss the constraints from the currently available LHC data, and then briefly mention implications for the future collider phenomenology.

%%%%%

\subsection{Current LHC limits on the technipions}

\subsubsection{Color-octet technipion $\theta_a^0$}

\begin{table} 
\begin{tabular}{c|c|c|c|c|c|c} 
\hline  \hline 
\hspace{10pt}
$N_{\rm TC}$ 
\hspace{10pt} &
\hspace{10pt} 
$m_{\theta_a^0}$ [TeV]
\hspace{10pt} 
&
\hspace{10pt}
$ \Gamma^{\rm tot}_{m_{\theta_a^0}}/m_{\theta_a^0} $  
\hspace{10pt}& 
\hspace{10pt}
${\rm BR}_{gg}$ (\%) 
\hspace{10pt}
 & 
\hspace{10pt}
 ${\rm BR}_{gZ,g\gamma}$ (\%) \hspace{10pt}
 & 
\hspace{10pt}
 ${\rm BR}_{tt}$ (\%) \hspace{10pt}
 & 
\hspace{10pt}
 ${\rm BR}_{bb, cc}$ (\%)  \hspace{10pt}
 \\ 
\hline 
3 & (4.6, 2.8, 1.6) & (0.15, 0.064, 0.030) & (49, 26, 11) & $\sim $ 0 & (51, 74, 89) & $\sim$ 0 \\ 
4 & (4.4, 2.7, 1.5) & (0.20, 0.075, 0.031) & (61, 37, 16) & $\sim $ 0 & (39, 63, 84) & $\sim$ 0 \\ 
5 & (4.3, 2.6, 1.5) & (0.26, 0.086, 0.034) &  (70, 46, 23) & $\sim$ 0 & (30, 54, 77) & $\sim$ 0 \\ 
\hline \hline 
\end{tabular} 
\caption{
The decay properties of the color-octet technipion $\theta_a^0$ (total width $\Gamma^{\rm tot}_{m_{\theta^0_a}}$ normalized to the mass and branching ratios, BR)  
for $N_{\rm TC}=3, 4$ and 5 with the predicted masses ($m_8$) listed in Table~\ref{tab:m38:vals}.   
}
\label{tab:38:decay} 
\end{table}

\begin{table} 
\begin{tabular}{c|c|c|c} 
\hline \hline 
\hspace{15pt} 
$N_{\rm TC} $
\hspace{15pt} 
& 
\hspace{15pt} 
$m_{\theta_a^0}\,[{\rm TeV}] $
\hspace{15pt} 
& 
\hspace{15pt} 
$\sigma_{\rm ggF}^{\theta_a^0} \times {\rm BR}(\theta_a^0 \to gg) \times {\cal A}_{\rm CMS} [{\rm fb}] $
\hspace{15pt} 
&  
\hspace{15pt} 
$\sigma_{\rm ggF}^{\theta_a^0} \times {\rm BR}(\theta_a^0 \to t \bar{t})  [{\rm fb}] $
\hspace{15pt} 
\\ 
\hline 
3 & (4.6, 2.8, 1.6) &(0.0004, 0.1, 2.2) & (0.011, 7.8, 510) \\ 
4 & (4.4, 2.7, 1.5) & (0.0019, 0.36, 8.4) & (0.034, 16, 1200) \\ 
5 & (4.3, 2.6, 1.5) & (0.0051, 0.96, 19) & (0.059, 30, 1700) \\ 
\hline \hline 
\end{tabular} 
\caption{ The predicted production cross sections times branching ratios of the color-octet technipion $\theta_a^0$ at the 8 TeV LHC for 
the reference values of masses 
listed in Table~\ref{tab:m38:vals} for $N_{\rm TC}=3,4,5$ and $S=(0.1,0.3,1.0)$. 
The production process is restricted to the gluon-gluon fusion (ggF), which is highly dominant at the LHC.     
${\cal A}_{\rm CMS}\simeq 0.6$ is the acceptance of digluon jets used at the CMS experiments~\cite{CMS-PAS-EXO-12-059}. 
} 
\label{gg:tt:events:theta} 
\end{table}

\begin{table} 
\begin{tabular}{c|c|c|c} 
\hline \hline  
\hspace{20pt} $m$ [TeV] \hspace{15pt} 
& 
\hspace{20pt} $\sigma_{gg}|_{\rm exp}^{\rm CMS8TeV}$ [fb] \hspace{20pt} 
& 
\hspace{20pt} $\sigma_{t\bar{t}}|_{\rm exp}^{\rm CMS8TeV}$ [fb] \hspace{20pt} 
& 
\hspace{20pt} $\sigma_{t\bar{t}}|_{\rm exp}^{\rm ATLAS8TeV}$ [fb] \hspace{20pt} 
\\ 
\hline 
(4.6, 2.8, 1.6) &  (0.94, 17, 150) & (---, 38, 120) &  (---, 74, 110) \\ 
(4.4, 2.7, 1.5) &  (1.1, 18, 120) &  (---, 40, 150) &  (---, 78, 140) \\ 
(4.3, 2.6, 1.5) &  (1.2, 22, 120) &  (---, 41, 150) &  (---, 75, 140) \\ 
\hline \hline 
\end{tabular} 
\caption{
The current 8 TeV LHC upper limits on the resonance mass $m$ from digluon jet and $t\bar{t}$ channels reported by the ATLAS and CMS collaborations read off from 
Refs.~\cite{CMS-PAS-EXO-12-059,ttbar}.  
The selected values of the resonance mass correspond to the $\theta_a^0$ mass predicted from the present holographic model displayed in Table~\ref{tab:m38:vals}.   
}  
\label{tab:limits:theta} 
\end{table}

The relevant cross sections and partial decay widths for two-body decays 
are read off from Ref.~\cite{Jia:2012kd}. 
 For the reference values for the mass of $\theta_a^0$  
given in Table~\ref{tab:m38:vals}, we compute the total width of $\theta_a^0$ and branching ratios to obtain the numbers shown in Table~\ref{tab:38:decay}.   
Table~\ref{tab:38:decay} implies that 
the digluon and $t\bar{t}$ events at the LHC may provide good probes for the discovery of $\theta_a^0$. 
In Table~\ref{gg:tt:events:theta} we show the predicted signal strengths of $\theta_a^0$ at the 8 TeV LHC for $gg$ and $t\bar{t}$ channels for 
the reference values of the $\theta_a^0$ masses listed in Table~\ref{tab:m38:vals}. 
 These signals can be constrained by the current LHC limits on searches for new resonances in the dijet and $t\bar{t}$ channels~\cite{CMS-PAS-EXO-12-059,ttbar} 
as shown in Table~\ref{tab:limits:theta}. 
Thus the current LHC data, especially on the $t \bar{t}$ channel, 
have already excluded the color-octet technipion $\theta_a^0$ with the mass $m_{\theta_a^0}\simeq 1.5$ 
for $(N_{\rm TC}, S)= (3, 1.0), (4, 1.0)$ and with the mass $\simeq 1.6$ TeV for $(N_{\rm TC}, S)=(5, 1.0)$.

\subsubsection{Color-singlet technipion $P^0$}

\begin{table}
\begin{tabular}{c|c|c|c|c|c|c} 
\hline \hline 
\hspace{8pt}  
$N_{\rm TC}$ 
\hspace{8pt} 
&
\hspace{8pt}  
$m_{P^0}$ [TeV] 
\hspace{8pt} 
& 
\hspace{8pt}  
$ \Gamma^{\rm tot}_{m_{P^0}}/m_{P^0} $ 
\hspace{8pt}  
& 
\hspace{8pt}  
${\rm BR}_{gg}$ (\%)
\hspace{8pt}  
& 
\hspace{8pt}  
${\rm BR}_{\gamma\gamma, Z\gamma, ZZ}$ (\%)
\hspace{8pt}  
& 
\hspace{8pt}  
${\rm BR}_{tt}$ (\%)
\hspace{8pt}  
& 
\hspace{8pt}  
${\rm BR}_{bb, cc, \tau\tau}$ (\%)
\hspace{8pt}  
\\ 
\hline 
3 & (2.3, 1.3, 0.72) & (0.033, 0.024, 0.019)  & (43, 21, 8) & $\sim$ 0 & (57, 79, 92) & $\sim$ 0 \\ 
4 & (2.4, 1.4, 0.76) & (0.047, 0.028, 0.020)  & (59, 33, 14) & $\sim$ 0 & (41, 67, 87) & $\sim$ 0 \\ 
5 & (2.5, 1.4, 0.79) & (0.066, 0.035, 0.022) &  (71, 45, 21) & $\sim$ 0 &  (29, 56, 79) & $\sim$ 0 \\ 
\hline \hline 
\end{tabular}  
\caption{The decay properties of the color-singlet technipion $P^0$ 
(total width $\Gamma^{\rm tot}_{m_{P^0}}$ normalized to the mass and branching ratios, BR) 
for $N_{\rm TC}=3, 4$ and 5 with   
the predicted masses of $P^0$ listed in Table~\ref{tab:mP:vals}. 
}
\label{tab:P:decay} 
\end{table}

\begin{table} 
\begin{tabular}{c|c|c|c} 
\hline \hline 
\hspace{15pt}
$N_{\rm TC} $
\hspace{15pt}
& 
\hspace{15pt}
$m_{P^0} \, [{\rm TeV}] $
\hspace{15pt}
&
\hspace{15pt}
$\sigma_{\rm ggF}^{P^0} \times {\rm BR}(P^0 \to gg) \times {\cal A}_{\rm CMS} [{\rm fb}]$
\hspace{15pt}
&  
\hspace{15pt}
$\sigma_{\rm ggF}^{P^0} \times {\rm BR}(P^0 \to t \bar{t})  [{\rm fb}] $
\hspace{15pt}
\\ 
\hline 
3 & (2.3, 1.3, 0.72) & (0.10, 1.2, 5.6) & (3.4, 130, 1800) \\ 
4 & (2.4, 1.4, 0.76) & (0.20, 2.8, 15) & (3.0, 150, 2500) \\ 
5 & (2.5, 1.4, 0.79) & (0.23, 5.0, 31) & (2.5, 160, 3100) \\ 
\hline \hline 
\end{tabular} 
\caption{
The production cross sections times branching ratios for $P^0 \to gg$ and $t\bar{t}$ channels evaluated at the 8 TeV LHC for the $P^0$ mass listed in Table~\ref{tab:mP:vals}. 
The production process is limited to ggF, which is highly dominant at the LHC.     
The signal strengths for digluon events have been multiplied by ${\cal A}_{\rm CMS}\simeq 0.6$, 
the acceptance of digluon jets used at the CMS experiments~\cite{CMS-PAS-EXO-12-059}. 
}
\label{gg:tt:events:P0}
\end{table}

\begin{table} 
\begin{tabular}{c|c|c|c} 
\hline \hline  
\hspace{20pt} $m$ [TeV] \hspace{15pt} 
& 
\hspace{20pt} $\sigma_{gg}|_{\rm exp}^{\rm CMS8TeV}$ [fb] \hspace{20pt} 
& 
\hspace{20pt} $\sigma_{t\bar{t}}|_{\rm exp}^{\rm CMS8TeV}$ [fb] \hspace{20pt} 
& 
\hspace{20pt} $\sigma_{t\bar{t}}|_{\rm exp}^{\rm ATLAS8TeV}$ [fb] \hspace{20pt} 
\\ 
\hline 
(2.3, 1.3, 0.72) &  (44, 350, ---) & (48, 230, 660) &  (72, 170, 2300) \\ 
(2.4, 1.4, 0.76) &  (42, 210, ---) &  (45, 180, 580) &  (75, 160, 2100) \\ 
(2.5, 1.4, 0.79) &  (33, 210, ---) &  (45, 180, 580) &  (75, 160, 1600) \\ 
\hline \hline 
\end{tabular} 
\caption{
The current upper limits on the resonance mass $m$ from digluon jet and $t\bar{t}$ channels reported by ATLAS and CMS experiments at 8 TeV 
read off from Refs.~\cite{CMS-PAS-EXO-12-059,ttbar}.  
The selected values of the resonance mass correspond to the $P^0$ masses in Table~\ref{tab:m38:vals}.   
}  
\label{tab:limits:P0} 
\end{table}

The formulas for relevant cross sections and partial decay widths are read off from Ref.~\cite{Jia:2012kd}. 
 For the predicted values for the mass of $P^0$  
listed in Table~\ref{tab:mP:vals}, 
we calculate the total width of $P^0$ and branching ratios to obtain the values shown in Table~\ref{tab:P:decay}.  
 From this, one can see that the digluon and $t\bar{t}$ events at the LHC may be channels for discovery of $P^0$. 
 For each channel the predicted signal strengths of $P^0$ at the 8 TeV LHC for the reference values of the mass listed in Table~\ref{tab:mP:vals} are displayed in Table~\ref{gg:tt:events:P0}.  
In comparison with the current LHC limits listed in Table~\ref{tab:limits:P0}, 
we thus see that the color-singlet technipion $P^0$ has already been excluded for the masses 720, 760 and 790 GeV corresponding to the cases of  
$(N_{\rm TC}, S)=(3, 1.0), (4, 1.0), (5, 1.0)$, respectively.

\subsubsection{Color-triplet technipions $T_c^{0,i}$} 

The LHC experiments have placed stringent constraints on the 
scalar leptoquarks $({\rm LQ}_{1,2,3})$~\cite{LQ123}, in which the most stringent bound on the mass 
has been set on the second generation leptoquark ${\rm LQ}_2$ as 
\begin{equation} 
 m_{{\rm LQ}_2} \ge 1070 \, {\rm GeV} 
\,, \label{LQ-limit}
\end{equation}
  with 100\% branching ratio for the decay to $\mu \nu_\mu + 2j$  being assumed. 
Though the coupling property of the color-triplet technipion $T_c^{0,i}$ depends highly on modeling of the extended TC, 
we may apply the above strongest bound on the $T_c^{0,i}$ masses.  
Comparing the reference values of $T_c^{0,i}$ masses listed in Table~\ref{tab:m38:vals} with the bound in Eq.(\ref{LQ-limit}), 
we thus see that the current LHC data have excluded the color-triplet technipions $T_c^{0,i}$ with the masses at around 1.0 -- 1.1 TeV, corresponding to the cases of   
$(N_{\rm TC}, S)=(3,1.0), (4, 1.0), (5, 1.0)$.

%%%%
\subsection{Implications for technirho searches}
A typical signature of the dynamical EW symmetry breaking scenario at hadron colliders is the existence of the new vector particle, called the technirho, which is an analogue of the rho meson in QCD. In the case of one-family model, there are various kinds of technirho mesons from the viewpoint of the SM gauge representation, in a similar way as technipions have various SM gauge representations. In general, technirho mesons are produced through the mixing with the SM gauge boson which is produced by the Drell-Yan (DY) process, though how they decay depend on the mass relation among technirho mesons and technipions. The mass of technirho meson can be calculated by the holographic method as 
\begin{equation}
M_{\rho}  \simeq (3.6,\, 2.1,\, 1.1)  \ {\rm TeV}\ \ {\rm for}\ \ S = (0.1,\, 0.3,\, 1.0).
\end{equation}
($N_{\rm TC}$ dependence is negligible.) Due to the large enhancement of the technipion masses (see Tables~\ref{tab:mP:vals} and \ref{tab:m38:vals}), 
decay channels of technirho mesons to a pair of technipions are closed, therefore, they decay to SM particles or a SM particle plus 
one technipion. 
In the case of the iso-triplet technirho mesons ($\rho^{\pm, 3}$), 
they dominantly decay to a pair of gauge bosons or one gauge boson plus one technipion like 
\begin{eqnarray} 
pp & \stackrel{\rm DY}{\to} &  \rho^\pm \to W^\pm + P^3  \quad {\rm or} \quad  Z + P^\pm 
\,, \nonumber \\ 
pp & \stackrel{\rm DY}{\to} &  \rho^3 \to W^\pm + P^\mp 
\,.  
\end{eqnarray}
In the case of the iso-singlet and color-singlet ($\rho^0$)/color-octet technirho mesons ($\rho_8^0$), 
there is a possibility that those technirhos dominantly decay to the technidilaton $(\phi)$ and the photon/gluon like 
\begin{eqnarray} 
 pp & \stackrel{\rm DY}{\to} & \rho^0 \to \phi + \gamma  
\,, \nonumber \\ 
 pp & \stackrel{\rm DY}{\to} & \rho^0_8 \to \phi + g 
\,.  
\end{eqnarray} 
The detailed study of collider phenomenologies will be presented in the future publications.

%%%%%
\section{Summary}
\label{sec:summary}
In this paper, we calculated the masses of the technipions in the one-family WTC model based on a holographic approach, 
which is known to be successful in the case of QCD. 
It was shown that technipion masses are enhanced due to the walking dynamics 
to 
as large as 
${\cal O} (1)$ TeV, somewhat larger than the previous estimates~\cite{Jia:2012kd}. 
Constraints from the currently available  LHC data, as well as implications for the future collider phenomenology were also discussed. 
In particular, we found a correlation between the technipion masses and 
the technicolor contribution to the $S$ parameter, $S^{({\rm TC})}$, which gives a constraint on the 
WTC model building solely from the current LHC data on the technipion mass limit: 
$S^{({\rm TC})}$ should not  be 
as large as ${\cal O} (1)$. 
This is a new constraint on the contribution to the $S$ parameter from the technicolor dynamics alone, in contrast to the  full $S$ parameter constraint from the precision electroweak measurements.

%%%%%
\acknowledgements
We would like to thank Koji Terashi for useful information and discussions.
This work was supported by 
the JSPS Grant-in-Aid for Scientific Research (S) \#22224003 and (C) \#23540300 (K.Y.). 

%%%%%
\appendix

%%%%%
\section{$\pi^+ - \pi^0$ mass difference in the real-life QCD} 
\label{app:pion}

\begin{figure}[t] 
\begin{center} 
\includegraphics[width=9cm]{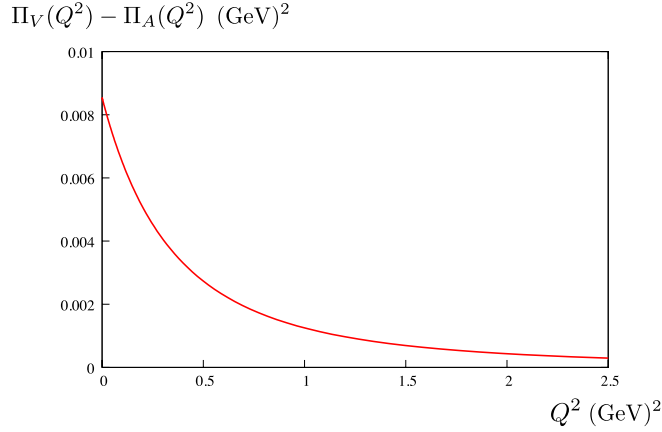} 
\end{center}
\caption{$\Pi_V(Q^2)-\Pi_A(Q^2)$ in the real-life QCD obtained from the holographic calculation with the input parameters shown in Eq.~(\ref{eq:QCDparam}).}  
\label{fig:QCD}
\end{figure}

Here, we apply the same method used in subsection~\ref{subsec:coloredTP} for the calculation of the $\pi^+ - \pi^0$ mass difference in the real-life QCD. The dominant part of the mass difference between $\pi^+$ and $\pi^0$ comes from the explicit breaking of the chiral symmetry due to the electromagnetic interaction. The formula for this mass difference is quite similar to the one in Eq.~(\ref{m38eq}): we just have to replace $\alpha_s(Q^2) \rightarrow \alpha_{\rm EM} \simeq 1/137$, $C_{3,8} \rightarrow 1$, and identify $F_\pi$ as the pion decay constant $f_\pi \simeq 92$ MeV:
\begin{eqnarray} 
 \Delta m_\pi^2 & \equiv & 
m_{\pi^+}^2-m_{\pi^0}^2 
\nonumber \\ 
&=& \frac{3\, \alpha_{\rm EM}}{4\pi f_\pi^2} \int_0^\infty dQ^2  \left[ \Pi_V(Q^2) - \Pi_A(Q^2) \right] 
\,. 
\label{eq:massdiff}
\end{eqnarray} 
Parameters of the holographic model are chosen so that it reproduces correct asymptotic behavior of the current correlators and experimental values of physical quantities. The optimal choice of the parameters is \cite{Haba:2010hu}:
\begin{equation}
 \gamma_m = 0,\ \ \ \xi=3.1,\ \ \ G=0.25,\ \ \ z_m^{-1}=347\ {\rm MeV}.
\label{eq:QCDparam}
\end{equation}
$\Pi_V(Q^2) - \Pi_A(Q^2)$ obtained from the holographic calculation with the above input parameters is shown in Fig.~\ref{fig:QCD}.
By inserting this $\Pi_V(Q^2) - \Pi_A(Q^2)$ into Eq.~(\ref{eq:massdiff}), we obtain $\Delta m_\pi^2 \simeq (32\ {\rm MeV})^2$, 
in reasonable agreement with experimental value $\Delta m_\pi^2 \simeq (35\ {\rm MeV})^2$~\cite{Beringer:1900zz}. 
We have also calculated $\Delta m_\pi^2$ with taking $G=0$ so that we could see the effect of gluon condensation, though the result was almost the same as in the case of $G=0.25$. This is reasonable since gluon condensation effect is expected to contribute to $\Pi_{V}(Q^2)$ and $\Pi_{A}(Q^2)$ in the same way, so the effect cancels in the integration of Eq.~(\ref{eq:massdiff}). The result can be translated to the form $\Delta m_\pi \equiv m_{\pi^+}-m_{\pi^0} = \frac{\Delta m_\pi^2}{m_{\pi^+}+m_{\pi^0}} \simeq 3.7\ {\rm MeV}$, where we used experimental values of $m_{\pi^+}+m_{\pi^0}$ in the denominator. This is compared to the result in Ref.~\cite{DaRold:2005zs} obtained by using the same holographic model except that, in their calculation, no gluonic-condensation effect is incorporated. They obtained $\Delta m_\pi \simeq 3.6 - 4.0\ {\rm MeV}$ depending on the choice of input parameter $\xi$, which is in reasonable agreement with our result.

%%%%%
\section{Comparison to the ladder BS calculation}
\label{app:ladder}
\begin{figure}[t] 
\begin{center} 
\includegraphics[width=9cm]{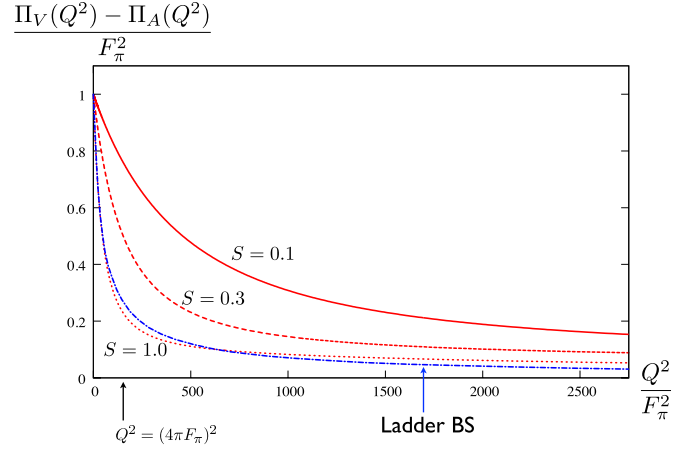} 
\end{center}
\caption{$\Pi_V(Q^2)-\Pi_A(Q^2)$ calculated from the holographic WTC model for the case of $N_{\rm TC} = 3$, compared with that calculated from the inhomogeneous BS equation with the improved ladder approximation in Ref.~\cite{Harada:2005ru}. Solid, dashed, and dotted (red) curves correspond to the cases of $S=0.1$, $0.3$, and $1.0$ obtained from the holographic calculation, while dashed-dotted (blue) curve is the one calculated from ladder BS equation. Both the vertical and the horizontal axes are normalized by $F_\pi^2$}  
\label{fig:comp}
\end{figure}

In this appendix, we compare $\Pi_V(Q^2)-\Pi_A(Q^2)$ calculated from the holographic WTC model with that calculated from the inhomogeneous BS equation with the improved ladder approximation in Ref.~\cite{Harada:2005ru}. In Fig.~\ref{fig:comp}, we show $\Pi_V(Q^2)-\Pi_A(Q^2)$ for the case of $N_{\rm TC} = 3$ (solid, dashed, and dotted (red) curves correspond to the cases of $S=0.1$, $0.3$, and $1.0$) obtained from the holographic calculation, along with the one calculated from ladder BS equation (dashed-dotted (blue) curve). Both the vertical and the horizontal axes are normalized by $F_\pi^2$. 
$\Pi_V(Q^2)-\Pi_A(Q^2)$ obtained from ladder BS calculation shown in the figure is same as that shown in Fig. 6 in Ref.~\cite{Harada:2005ru}. See Ref.~\cite{Harada:2005ru} for detailed explanation for the calculation. From the figure, we see that the one calculated from the ladder BS equation is similar to the one calculated from holographic method with $S=1.0$, and each gives the contribution to the mass of the color-triplet technipion 1.2 and 1.1 TeV, respectively. This similarity can be understood from the fact that the ladder BS calculation show that the contribution to the $S$ parameter from one EW doublet (denoted as $\hat{S}$) is about $0.3$~\cite{Harada:2005ru}, while $S=1.0$ in one-family model means $\hat{S}=0.25$ since one-family model has four EW doublet. The slope of $\Pi_V(Q^2)-\Pi_A(Q^2)$ is proportional to the magnitude of $\hat{S}$, thus the similarity between ladder calculation and the holographic calculation with $S=1.0$ ($\hat{S}=0.25$) is a confirmation of consistency between two calculation methods.

%%%%%

\end{document}